\documentclass[final,3p,times,twocolumn]{elsarticle}

 \usepackage{graphics}

\usepackage{amssymb}

\usepackage{lineno}

\usepackage{subfigure}

\journal{Nuclear Instruments and Methods A }

\begin{document}

\begin{frontmatter}




\title{Vertex Tracking at a Future Linear Collider}
\author{M.~Battaglia}
\ead{MBattaglia@lbl.gov}
\address{Santa Cruz Institute of Particle Physics, University of California 
at Santa Cruz, Santa Cruz CA 95064, USA\\
Lawrence Berkeley National Laboratory, Berkeley CA 94720, USA\\
CERN, Geneva CH-1211 Switzerland}

\begin{abstract}
The anticipated physics program at an high energy $e^+e^-$ linear collider 
places special emphasis on the accuracy in extrapolating charged particle 
tracks to their production vertex to tag heavy quarks and leptons. This 
paper reviews physics motivations and performance requirements, sensor 
R\&D directions and current results of the studies for a vertex tracker 
at a future linear collider. 
\end{abstract}

\begin{keyword}
Pixel detectors  \sep linear colliders 

\end{keyword}

\end{frontmatter}


\section{Introduction}
\label{sec:1}
A high energy $e^+e^-$ linear collider has emerged as possibly the most practical and 
realistic way towards collisions of elementary particles at constituent energies matching 
those of the LHC with high luminosity. The ILC 
project is based on the use of superconducting RF cavities providing gradients of 
$\simeq$~30~MV/m to produce collisions at centre-of-mass energies $\sqrt{s}$ = 0.25 - 
1~TeV~\cite{Brau:2007zza}. In order to achieve multi-TeV $e^+e^-$ collisions, the 
CLIC project develops a new acceleration scheme where a low-energy, high-current drive 
beam is used to accelerate the main beam through high-frequency transfer structures, which 
have achieved gradients in excess of 100 MV/m~\cite{Assmann:2000hg}.

Heavy flavours represent an essential signature of the anticipated physics of interest. 
The study of the Higgs sector of the Standard Model, of TeV-scale new physics and the search 
for new phenomena at very high mass scale through electro-weak precision observables, all depend 
on the identification and decay reconstruction of $t$, $b$ and $c$ quarks and of $\tau$ 
leptons~\cite{beach2010}. This physics program requires a vertex tracker able to extrapolate 
the particle tracks back to their production vertex with high accuracy over a broad momentum 
range. The linear collider requirements have motivated a vigorous and diversified R\&D program
which has seen monolithic pixels of various technologies emerging as a mature and well-performing 
option for vertex tracking applications. 

\section{Tracking Accuracy, Flavour Tagging and Experimental Conditions}

The standard figure of merit for tracking accuracy is the resolution on the impact parameter, 
$\sigma_{IP}$, defined as the distance of closest approach of the particle track to the colliding 
beam position. This can be parametrised as:
\begin{equation}
\sigma_{IP} = a \oplus \frac{b}{p sin^{k} \theta} 
\label{eq:ip}
\end{equation}
where $\theta$ is the track polar angle and $k$ = 3/2 for the $R-\Phi$ and 5/2 for the $z$ projection.
The target parameters for a 0.25-0.5~TeV collider are $a$ = 5~$\mu$m and $b$ = 10~$\mu$m~GeV$^{-1}$.

The identification of hadronic jets originating from heavy quarks is best achieved by a 
topological reconstruction of the displaced secondary and tertiary vertex structure and the 
kinematics associated to $B$ hadron decays. The ability to reconstruct the sequence of primary, 
secondary and tertiary vertices depends on the impact parameter resolution.
Jet flavour tagging for the linear collider extends the strategy successfully adopted in SLD, 
to date the collider experiments with the best track extrapolation accuracy~\cite{Abe:1999ky}. 
The ZVTOP algorithm~\cite{Jackson:1997}, originally developed for physics at SLC and now adapted 
for use at a linear collider, demonstrated high $b$-tagging efficiency on fully simulated and 
reconstructed events~\cite{Bailey:2009ui}. 
The impact of a change of the $a$ and $b$ parameters in~(\ref{eq:ip}) on the physics performance  
have been studied on detailed simulation and reconstruction. 
In 500~GeV energy $b$-jets, doubling the values of $a$ and $b$ from 3~$\mu$m and 
18~$\mu$m~GeV$^{-1}$ to 6~$\mu$m and 36~$\mu$m~GeV$^{-1}$, respectively, results in a 15~\% decrease 
of both the number of vertices reconstructed in the $B$ decay chain and the fraction of particle tracks 
correctly assigned to their vertex of origin. The efficiency for the identification of $b$ jets at a 
constant purity of 0.90, in a sample where light, $c$ and $b$ flavours are uniformly represented, 
drops from 0.75 for $a$ = 5~$\mu$m and $b$ = 10~$\mu$m~GeV$^{-1}$ to 0.25 for $a$ = 12~$\mu$m and 
$b$ = 70~$\mu$m~GeV$^{-1}$. That for tagging $c$-jets at a purity of 0.70 drops from 0.50 for  $a$ = 
5~$\mu$m and $b$ = 10~$\mu$m~GeV$^{-1}$ to 0.29 for $a$ = 11~$\mu$m and $b$ = 
15~$\mu$m~GeV$^{-1}$~\cite{hawkings}. 
Propagating these effects to the statistical accuracy of physics measurements, such as Higgs decay 
branching fractions, generally shows that a degradation of a factor of two on the $a$ or $b$ terms 
of~(\ref{eq:ip}) corresponds to a 20-30\% equivalent luminosity loss at 0.5~TeV~\cite{kuhl,yu}.  

Multiple $t$ and $b$ quarks are expected to be a distinctive feature in several processes
in multi-TeV $e^+e^-$ collisions. The signal cross sections are typically of 
$\cal{O}\mathrm{(1~fb)}$ with signal-to-background ratios of $10^{-2}$-$10^{-5}$ and two to 
six heavy flavour jets. The energy of $b$ jets ranges from 50~GeV up to the full beam energy 
of $\sim$1.5~TeV. 
In the benchmark process $e^+e^- \to H^0A^0 \to b \bar b b \bar b$, with $M_A$ around 1~TeV at 
$\sqrt{s}$=3~TeV,  jets have $B$ hadrons flying on average 27~mm and approximately one third of them 
decays at a radius larger than 30~mm. In addition, forward $b$-tagging is crucial to study some 
specific SM processes, 
which may only be accessible to a multi-TeV $e^+e^-$ collider, such as the triple Higgs coupling, 
through $e^+e^- \to \nu_e \bar \nu_e h^0 h^0$, and the fermionic coupling of an intermediate-mass 
Higgs boson, through $e^+e^- \to \nu_e \bar \nu_e h^0 \to \nu_e \bar \nu_e b \bar b$. In this case,
the $B$ hadron energy is in the range 25 $< E < $ 300~GeV, due to the large energy taken by the 
neutrinos.

The track extrapolation requirements outlined above can be met with a vertex tracker which has thin 
layers, first measurement close to the beam interaction point and excellent single point resolution. 
Considering a traditional barrel geometry with two layers located at radii 
$R_{in}$ and $R_{out}$, the asymptotic impact parameter resolution $a$ is given by:
\begin{equation} 
a = \sqrt{(n+1)^2 + n^2} \sigma_{point}
\end{equation} 
where $n = \frac{R_{in}}{R_{out}-R_{in}}$ and $\sigma_{point}$ is the single point resolution. 
With $R_{in} =$ 15-30~mm and a lever arm 
$R_{out}-R_{in}$ of 30 - 40~mm, the requirement $a \le$ 5~$\mu$m implies a point resolution
$\sigma_{point} \le$ 3.5~$\mu$m.
The requirement on the pixel pitch, $P$, comes from the point resolution 
($\sigma_{point} \le P/\sqrt{12}$ for binary readout and $\propto P/(S/N)$ for analog readout with 
charge interpolation) but also from the two-track separation ($\propto P$) and from the occupancy 
($\propto P^2$). Single point resolutions equal to, or better than, that required for the linear collider 
have been obtained in several monolithic pixel technologies, including CMOS MAPS~\cite{winter}, 
DEPFET~\cite{depfet-beam} and SOI~\cite{soi-pixel2010}, for pixel pitches in the range 10-30~$\mu$m and 
S/N values of 20 to 130. In particular, CMOS sensors with binary output have demonstrated a detection 
efficiency $>$99~\%, fake rate $<$10$^{-4}$ and point resolution of 3.5~$\mu$m, below the $P/\sqrt{12}$ limit 
of their pixel pitch of 18.4~$\mu$m~\cite{winter,demasi}.

The multiple scattering term, $b$, is given by:
\begin{equation} 
b = \sum_{i} R_i^2 \theta_{0,i}^2
\end{equation} 
where the index $i$ runs over the material surfaces traversed by the particles and $\theta_0$ is the multiple 
scattering angle for normal incidence tracks. It is desirable to install the first layer as close as possible 
to the interaction point to minimise the multiple scattering effect and optimise the asymptotic resolution. 
In practice this radius is set by the density of low momentum pair electrons and positrons. 

Pairs are produced by beam particles scattering on real and virtual photons created in the intense beam-beam 
electromagnetic interactions. 
The intrinsic $p_t$ of most of these electrons is small but they get deflected in the electric field 
of the incoming beam and can reach the detector. Incoherent pairs, spiralling in the solenoidal 
magnetic field of the detector, form an envelope, whose radius at a given position $z$ along the beam 
axis defines a lower bound on the position of the innermost detector layer. This bound 
scales approximately as:
\begin{equation} 
R \sim \sqrt{\frac{N}{10^{10}} \frac{1}{\sigma_z} \frac{1}{B} z}.
\end{equation} 
$N$ is the number of particles in a bunch (= 2$\times$10$^{10}$ for ILC and 3.7$\times$10$^{9}$ for 
CLIC), $\sigma_z$ 
is the bunch length (= 300~$\mu$m for ILC and 40~$\mu$m for CLIC) and $B$ the solenoidal field (= 3-5~T 
depending on the detector concept design)~\cite{scaling,scaling2}. Outside the deflected pair envelope 
there is a residual pair population, due to large $p_t$ electrons and to particles deflected at large angles 
or back-scattered from the inner face of the low angle calorimeter, which is located downstream from the 
interaction point. The ILC has a predicted hit density from incoherent pairs of 4.4~hits mm$^{-2}$ BX$^{-1}$ 
at a radius of 16~mm and $\sqrt{s}$ = 0.5~TeV, which increases by approximately a factor of two
by doubling the beam energy or changing the beam parameters to achieve a constant luminosity. 
At $\sqrt{s}$ = 3~TeV, CLIC has $\simeq$2.2~hits mm$^{-2}$ BX$^{-1}$ at a radius of 31~mm~\cite{clic-pairs}. 
The factor two increase in the radius of the vertex tracker innermost layer at CLIC compared 
to the ILC results in a change of the multiple scattering term from 10~$\mu$m GeV$^{-1}$ to 
21~$\mu$m GeV$^{-1}$.
Pairs are also responsible for an ionising dose of $\sim$100~krad/year to be added to a non-ionising 
dose corresponding to $\simeq$ 7$\times$10$^{10}$ n$_{eq}$ cm$^{-2}$ year$^{-1}$ from pairs, which is 
larger than that due to neutrons, estimated to $\simeq$ 10$^{10}$ neutrons cm$^{-2}$ year$^{-1}$ at 
0.5~TeV~\cite{besson}. These figures are expected to be comparable, $\sim$100~krad/year and 
5$\times$10$^{10}$ $n$ cm$^{-2}$ year$^{-1}$, for CLIC at 3~TeV~\cite{Battaglia:2004mw}. 

Experience in tracking and vertexing with prototype monolithic pixel sensors has already been obtained 
by several groups. In particular, the EUDET project and the DEPFET group have extensive experience with 
tracking beam particles. The EUDET telescope consists of two arms each equipped with three layers of 
50~$\mu$m-thin CMOS MAPS chips with in-pixel correlated double sampling, column parallel readout with 
discriminator and zero suppression logic at the end of the column. It provides a track extrapolation 
accuracy of 1-2~$\mu$m on the detector under test. Using a readout time of 100~$\mu$s it can operate with 
large track density, up to 10$^6$ particles cm$^{-2}$ s$^{-1}$~\cite{eudet-cmos}.
The T966 beam test experiment also operated a beam telescope made of 50~$\mu$m-thick 
CMOS pixel sensors~\cite{Battaglia:2008nj}. Four sensors were arranged with a 15~mm spacing which is close 
to that proposed for the ladders of a Vertex Tracker at the linear collider. It studied the tracking 
extrapolation accuracy for 1.5~GeV electrons at the LBNL Advanced Light Source (ALS) and 120~GeV protons 
at the Fermilab Test Beam Facility. With a point resolution of 2.3~$\mu$m, the accuracy for extrapolating 
the reconstructed particle track by 15~mm upstream was measured to be (8.5$\pm$0.4)~$\mu$m and 
(4.2$\pm$0.3)~$\mu$m at 1.5 and 120~GeV, respectively, which matches the linear collider requirements. 
In addition, the vertexing accuracy was studied for $p$ interactions in a thin Cu target located 32~mm 
upstream from the first sensor, corresponding to the distance between the first vertex layer 
and the interaction point foreseen at CLIC. The longitudinal vertex position resolution of 260~$\mu$m 
for an average track multiplicity of 2.74 closely matches that expected for CLIC of 220~$\mu$m 
for reconstructed secondary decays vertices of $B$ hadrons having an average track multiplicity of 3.02.

\section{Low-mass ladders for the Vertex Tracker}

Despite the large design collision energies, charged particles are 
typically produced with moderate energies, due to the large jet multiplicity or missing 
energy. $b$ jets are discriminated from $c$ jets based on the number and invariant mass of the 
secondary particles. This requires most, if not all, of the $b$ charged decay products to be 
identified. Excellent track extrapolation at low momenta is therefore 
essential. At 0.5~TeV the $c$ tagging efficiency, at constant purity, in the study of 
$h^0 \to c \bar c$,  drops by 25~\% when changing the ladder thickness from 0.1~\%~$X_0$ to
0.3~\%~$X_0$~\cite{Luzniak:2009}. The multiple scattering term $b$ plays even a more crucial 
role than the asymptotic resolution $a$, in particular for processes which are forward peaked. 
The fraction of charged $B$ decay products identified as secondaries based on their 
impact parameter significance, $IP/\sigma_{IP}$, in 
$e^+e^- \to \nu_e \bar \nu_e h^0 \to \nu_e \bar \nu_e b \bar b$ 
events at $\sqrt{s}$ = 3~TeV, drops from 0.85 for $b$ = 15~$\mu$m~GeV$^{-1}$ to 0.74 for 
35~$\mu$m~GeV$^{-1}$. For comparison, it changes by just 2~\% for 1.5 $< a <$ 3.5~$\mu$m.

Chips of various technologies have been successfully thinned to $\le$~100~$\mu$m. 
Assuming that the sensors are mounted 
on an 100$~\mu$m-thick carbon fiber composite (CFC) ladder, there is little benefit 
for the multiple scattering term from pushing their thickness below $\sim$50~$\mu$m. It has 
been shown that CMOS chips can be thinned to 50~$\mu$m with high yields and no loss in 
performance~\cite{Battaglia:2006aa}. For DEPFET sensors an interesting method of selective thinning 
of the handle wafer in the sensitive area while retaining a support frame for mechanical 
stability has been developed on large area prototypes and should be soon tested on sensor 
chips~\cite{depfet-thin}. 
Several designs of low material ladders have been proposed, based on specific sensor 
technologies. The ladders of the STAR HFT are probably the most advanced prototypes 
built so far. Based on CMOS MAPS thinned to 50~$\mu$m, the ladder has a total material 
budget of 0.37~\%~$X_0$, needed for the study of low-$p_t$ $D$ meson production in heavy 
ion collisions at RHIC~\cite{greiner-pixel2010}. The chosen geometry, with two barrel layers 
at 25 and 80~mm radius, will obtain a multiple scattering $b$ term of 19~$\mu$m~GeV$^{-1}$. 
In the DEPFET-based ladder proposed for Belle-II the material budget decreases to 0.19~\%~$X_0$ 
which becomes 0.17~\%~$X_0$ in the conceptual design of a FPCCD-based double-layer ladder envisaged 
for the ILC. It is interesting to observe that in both the STAR HFT and the Belle-II ladders the 
Si chip accounts for less than half of the ladder material budget. In particular, the kapton cable 
with Al traces installed below the sensors to route signals, power and clocks accounts  in the STAR 
design for 20~\% of the total material budget of the ladder. In order to progress towards thinner 
modules, sensor stitching, with clock and signals routed on metal lines in the chips, appears a 
promising path.

Different arrangements of detectors in a multi-layered barrel geometry are being evaluated.
The first is a traditional vertex tracker consisting of five equally-spaced layers with thin Si 
chips mounted on a light-weight carbon composite ladder support. Measurements carried out at LBNL with 
an optical survey machine on a free-standing 1.9$\times$1.9~cm$^2$ CMOS MAPS chip thinned to 
40~$\mu$m show a warping which is primarily cylindrical. The measured peak-to-peak amplitude is 
1.7~mm, which results in a significant stress when the chip is flattened on its carrier~\cite{lbnl}. 
The support structure must counter this deformation and ensure the planarity of the sensors on the 
ladder. An interesting solution is the use of a double-sided ladder made of a carbon foam layer with 
sensors mounted at both sides, in a sandwich-like structure. This design places the thin pixel 
chips back-to-back on the same carrier minimising ladder bending, thanks to the symmetric layout, 
and reducing the material budget per active layer~\cite{lbnl,ildloi}. It also provides 
hit position correlation on the closely-spaced detector planes, which may offer further advantages 
for redundancy and rejection of low-energy electrons from the pair background. The PLUME R\&D program 
foresees the development of light-weight double-layer ladder prototypes, based on CMOS sensors, 
with the aim to achieve a material budget of 0.2~\%~$X_0$~\cite{plume-pixel2010}.

The LHC detector experience shows that cooling and services dominate the total material budget. 
Passive cooling appears necessary to maintain the ladder material within few per-mil of a radiation
length. A simple air flow is adopted by the STAR HFT. Tests carried out at LBNL have shown that airflow 
at 10~m~s$^{-1}$ can remove up to $\simeq$200~mW~cm$^{-2}$ with an acceptable temperature rise 
above ambient~\cite{greiner-pixel2010,lbnl}. Results reported by the SiLC collaboration confirm 
this figure~\cite{silc}. This places stringent constraints on the chip power 
dissipation which must be reconciled with the requested data processing and readout speed. 
The MIMOSA-26 CMOS chip, realised in the framework of the EUDET-FP6 project and which represents a 
prototype for the sensor for the STAR HFT, has an average power dissipation of 280~mW~cm$^{-2}$~\cite{demasi}.
More advanced thermal management approaches, such as micro-channel cooling, are becoming more realistic 
following recent developments in the IC industry and are presently being evaluated~\cite{bosi}.

\section{Sensor Technologies and Architectures for the Vertex Tracker}

One of the most demanding requirements for the sensor architecture at a linear collider is the mitigation of 
the anticipated occupancy due to incoherent pairs, within the basic specifications of $\le$3.5~$\mu$m single 
point resolution, $\simeq$ 50~$\mu$m thickness, power dissipation $\le$~200~mW cm$^{-2}$ (assuming airflow 
cooling).  This can be achieved either by an 
architecture based on fast continuous readout or by in-situ charge storage with high space (or space-time) 
granularity and read-out at the end of the bunch train. At the ILC, which has 2820 bunches/train with a 
337~ns spacing and 5~Hz repetition rate, an occupancy low enough not to affect the pattern recognition 
should be achievable on the innermost layer by a readout within 25-50~$\mu$s. An example 
of continuous readout architecture is implemented on CMOS MAPS with double-sided, column-parallel binary 
read-out using a rolling shutter, correlated double sampling and zero suppression~\cite{demasi}. 
DEPFET pixels are read-out 20 times during the train, with data stored on the chip periphery~\cite{depfet-ilc}.
Alternatively, time-stamping can be performed to divide the bunch train into several time-buckets. 
Concepts currently under R\&D are the Chronopixels~\cite{chrono} and the 
ISIS pixels with in-situ 20-cell charge storage~\cite{isis, isis2}. Finally, small enough pixels can achieve 
a low occupancy even integrating over a full bunch train. The FPCCD concept is based on an extreme space 
granularity, where 5$\times$5~$\mu$m$^2$ CCD pixels ensure low occupancy and, possibly, background hit 
rejection through cluster shape analysis~\cite{fpccd}.

Continuous readout schemes must ensure immunity from pick-up during the bunch train and are 
challenging for background rates significantly higher than the current estimates for the ILC.
In-pixel time stamping, as pursued by the Chronopixel design, has to achieve noise reduction and high 
bandwidth within an acceptable power budget. The design targets a 45~nm CMOS process with deep implants 
to avoid competitive charge collection, still untested for pixel detectors. 
At an X-band multi-TeV collider, such as CLIC, both the bunch spacing (0.5~ns) and the train length 
(150~ns) are too short to achieve a reduction of the integrated background by fast readout. 
Instead, high space-time granularity, such as time-stamping to 15-30~ns, must be considered, 
given the expected background rate of $\sim$5~hits~mm$^{-2}$ per bunch train.
Integrating fast time stamping in a thin sensor, with small pixel cells and limited power 
consumption requires significant R\&D, which can profit of features made available by 
emerging semiconductor technologies.

The R\&D driven by the ILC specifications has made significant progress in the last decade 
towards sensors which are now ready for applications in real experiments (STAR, CBM, Belle-II) 
with specifications which are in many areas close to those of a linear collider. 
At STAR, where there are $\simeq$2~particles mm$^{-2}$ in the 200~$\mu$s integration time, 
sensors with binary output and on-chip zero suppression are foreseen for installation in 2013.
The expected ionising does is $\sim$100~krad/year. At Belle-II the hit density is expected to be 
$\simeq$8~hits mm$^{-2}$ in the 20~$\mu$s integration time of a DEPFET detector, which makes it 
an even bigger challenge compared to a linear collider. The ionising dose should exceed 1~Mrad/year.
The construction and operation of these detectors will also provide a valuable experience with 
those aspects which have been studied less so far, such as yields, system issues and operation 
reliability.

Beyond particle physics, new applications in imaging and X-ray detection have been made possible 
by the linear collider R\&D. Imaging in transmission electron microscopy (TEM) with CMOS 
sensors is under fast development and profits from the availability of thin sensors with thin 
sensitive volume to minimise low energy electron scattering and thus the point spread 
function~\cite{emicro, deptuch, team}. 
Given that a very intense bright field image could deposit order of 10~rad~s$^{-1}$ pixel$^{-1}$, a 
target radiation tolerance of $\ge$~1~Mrad is a valid requirement. The development of sensors for 
TEM applications has contributed useful results in terms of design and tests of radiation-tolerant 
pixel cells, able to operate after doses of several Mrad~\cite{team}. The integration of CMOS sensors 
into photosensitive systems for fluorescent and bio-luminescent high-speed imaging in biology has 
already provided a characterisation of a first prototype with different fluorescent dyes~\cite{barbier}. 
Low noise and high energy resolution make DEPFET sensors ideal for X-ray spectroscopy at high rates. 
Applications both at the European XFEL~\cite{depfet-xfel} and for the focal plane of a planetary 
observation satellite experiment~\cite{depfet-mer} are presently being developed.

Despite the successes in the development of a new generation of pixel sensors for application at a
linear collider, there are still several areas where more sensor R\&D is needed. Fast charge collection 
and signals larger than the $\le$10$^3$ electrons generated in the thin epitaxial layer of conventional 
bulk CMOS processes are highly desirable to ensure radiation tolerant devices and the possibility
of fast time stamping. The integration of not only signal sensing but also advanced data processing 
in pixel, or at least on chip, will also profit from the introduction of new technologies. The port 
of MAPS to commercially available processes with an high resistivity epitaxial layer or substrate 
is actively pursued. These offer higher charge yield, faster charge collection dominated by drift 
and improved radiation tolerance compared to conventional bulk CMOS technology. MIMOSA sensors with 
high-resistivity epi-layer have been successfully tested and will be used in the STAR HFT~\cite{winter}. 
CMOS pixel sensors with an high-resistivity sensitive volume are being developed by the LEPIX collaboration 
centred at CERN in a custom 90~nm process. The first structures are presently being evaluated. 
The Silicon-On-Insulator 
process with an high-resistivity handle wafer offers further appealing opportunities by removing limitations 
of bulk CMOS processes. After the pioneering effort in a 3~$\mu$m CMOS process of~\cite{krakow}, a commercial 
0.2~$\mu$m SOI process has been made available since a few years by a collaboration of KEK with OKI 
Semiconductors, Japan, ~\cite{soi}. The high resistivity sensitive volume ensures large signals with no 
interconnections, low collection electrode capacitance and a low-power, potentially radiation tolerant 
device~\cite{Battaglia:2007eq,soi2}. The R\&D has successfully solved the transistor back-gating problem. 
Prototype SOI pixels recently tested with a 200~GeV $\pi^{-}$ beam at CERN have been successfully operated 
with a depletion voltage up to 70~V, giving a preliminary result of (1.2$\pm$0.05)~$\mu$m single point 
resolution and particle detection efficiency above 98~\%~\cite{soi-pixel2010}. Similar features, but 
allowing also the integration of heterogeneous CMOS technologies can be obtained using 3D multi-tier 
vertical integration of thin Si chips. Several pilot R\&D programs have recently been launched to explore 
the use of 3D inter-connectivity for pixel sensors and results are expected soon~\cite{soi2,yarema}.

\section{Conclusions}
\label{sec:conclusions}

Physics requirements at a future linear collider have motivated a significant R\&D effort on 
monolithic pixel sensors to achieve small pixel cells with integrated charge sensing and data 
processing, thin sensors with low power consumption and high space-time granularity.
Current studies show that moving from a sub-TeV to a multi-TeV collider implies new requirements 
especially on forward vertexing capabilities and detector fast time stamping, which need to be 
addressed by a dedicated R\&D program. The LHC data will likely indicate the energy range of 
interest for a linear collider in the next few years. Despite the long time scale of these programs, 
technologies developed in the ILC-motivated R\&D have already demonstrated a significant impact on 
other particle physics experiments as well as imaging and spectroscopy applications in other fields 
of science from electron microscopy to biology and astronomy.

\section{Acknowledgement}
\label{sec:acknowledgment}
I am grateful to the organisers of the PIXEL 2010 workshop for their invitation and an enjoyable 
conference and to L.~Andricek, F.~Bogert, J.~Brau, D.~Contarato, R.~De~Masi, P.~Denes, P.~Giubilato, 
L.~Greiner, A.~Sailer, D.B.~Shuman, W.~Snoeys and Y.~Sugimoto for their contributions.

\end{document}